# Evolution of morphological and physical properties of laboratory interstellar organic residues with ultraviolet irradiation

*L. Piani[1*], S. Tachibana[1], T. Hama[2], H. Tanaka[2,3], Y. Endo[1], I. Sugawara[1], L. Dessimoulie[1], Y. Kimura[2], A. Miyake[4], J. Matsuno[4], A. Tsuchiyama[4], K. Fujita[2], S. Nakatsubo[2], H. Fukushi[2], S. Mori[2], T. Chigai[2], H. Yurimoto[1] and A. Kouchi[2]*

[1]*Department of Natural History Sciences, Science Faculty, Hokkaido University, Sapporo 060-0810, Japan.*

[2]*Institute of Low Temperature Science, Hokkaido University, Sapporo 060–0819, Japan.*

[3]*Astronomical Institute, Tohoku University, Sendai 980-8578, Japan*

[4]*Division of Earth and Planetary Science, Kyoto University, Kyoto 606-8502, Japan.*

*corresponding author: laurette@ep.sci.hokudai.ac.jp

**Short title for running head**

UV-irradiation of molecular cloud organics

**Abstract**

Refractory organic compounds formed in molecular clouds are among the building blocks of the Solar System objects and could be the precursors of organic matter found in primitive meteorites and cometary materials. However, little is known about the evolutionary pathways of molecular cloud organics from dense molecular clouds to planetary systems. In this study, we focus on the evolution of the morphological and viscoelastic properties of molecular cloud refractory organic matter. We found that the organic residue, experimentally synthesized at ~10 K from UV-irradiated $H_2O$-$CH_3OH$-$NH_3$ ice, changed significantly in terms of its nanometer- to micrometer-scale morphology and viscoelastic properties after UV irradiation at room temperature. The dose of this irradiation was equivalent to that experienced after short residence in diffuse clouds ($\leq 10^4$ yrs) or irradiation in outer protoplanetary disks. The irradiated organic residues became highly porous and more rigid and formed amorphous nanospherules. These nanospherules are morphologically similar to organic nanoglobules observed in the least-altered chondrites, chondritic porous IDPs, and cometary samples, suggesting that irradiation of refractory organics could be a possible formation pathway for such nanoglobules. The storage modulus (elasticity) of photo-irradiated organic residues is ~100 MPa irrespective of vibrational frequency, a value which is lower than the storage moduli of minerals and ice. Dust grains coated with such irradiated organics would therefore stick together efficiently, but growth to larger grains might be suppressed due to an increase in aggregate brittleness caused by the strong connections between grains.

## 1. Introduction

Interstellar molecular clouds, the birthplace of stars, are composed of gases and dust grains (silicates and carbonaceous materials). These low-temperature environments are favorable places for condensation of ice mantles, mainly composed of $H_2O$, $CO$, $CO_2$, $CH_3OH$, and $NH_3$, on the dust grains (e.g. Dartois 2005 and reference therein). Grain-surface reactions, direct energy deposition by cosmic-rays impinging on dust grains and secondary-ultraviolet (UV) photons generated by cosmic-ray excitation of hydrogen, and thermal heating can all modify the ice mantles, resulting in the formation of more complex and refractory organic components.

A number of laboratory experiments have been carried out to study the formation and evolution of the ice and organic molecules under interstellar medium (ISM) conditions (e.g. Greenberg 2002 and references therein). In the laboratory, thin ice films produced by the condensation of a mixture of simple gas species on a cooled substrate (~10 K) are irradiated by UV photons or ions after or during gas deposition. During heating of the ice to room temperature, radicals formed by irradiation react to produce complex organic molecules. The refractory organic residue that remains on the substrate, sometimes referred as 'yellow stuff', has been considered an analogue of the organic matter formed in the dense molecular cloud (Greenberg 1983; Agarwal et al. 1985; Jenniskens et al. 1993; Bernstein et al. 1995; Muñoz-Caro and Schutte 2003). This residue is hereafter referred to as MCOR, for Molecular Cloud Organic Residue, in accordance with previous studies (e.g. Kouchi et al. 2002; Kudo et al. 2002). The organic residue can be further irradiated to simulate conditions in the diffuse cloud, where both the UV photon flux and gas temperature are higher than in the dense molecular cloud (Jenniskens et al. 1993). These conditions might also reflect to the irradiation of organic compounds in the outer solar nebula (Chiang and Godreich 1997; Ciesla and Sandford 2012). The remaining residue is hereafter referred to as DCOR, for Diffuse Cloud Organic Residue (Kouchi et al. 2002; Kudo et al. 2002).

Organic compounds in chondrites, interplanetary dust particles (IDPs), and comets might have been inherited from the molecular cloud chemistry (Mumma and Charnley 2011; Flynn et al. 2013; Bockelée-Morvan et al. 2015; Goesmann et al. 2015). Macromolecular organic materials with deuterium (D) and/or $^{15}$N-enrichments in these extraterrestrial samples are thought to have recorded a low-temperature chemistry possibly occurring in the Sun's parent molecular cloud (Floss et al. 2004; Busemann et al. 2006; Nakamura-Messenger et al. 2006; Hashiguchi et al. 2015), however details of their formation and evolution remain unclear (Ceccarelli et al. 2014).

Morphological variations in the macromolecular organic matter of extraterrestrial objects indicate their complex evolution and argue for multiple origins. For instance, organic nano-globules in chondrites (Nakamura et al. 2002; Garvie and Buseck 2004), IDPs (Flynn et al. 2003), and comet particles (De Gregorio et al. 2010) show textural variations at micrometer and nanometer scales. Ring globules, round globules, irregular-shaped globules, and globule aggregates are ubiquitous in the matrices of carbonaceous chondrites, often bearing D-rich anomalies and sometimes associated with silicates or oxides (Hashiguchi et al. 2013). Different formation processes have been suggested such as organic coating on pre-existing grains (Nakamura-Messenger et al. 2006) and formation by gas ionization in high temperature plasmas (Saito and Kimura 2009) or during fluid-assisted parent body alteration (Cody et al. 2011), but the details of the formation pathways remain puzzling (De Gregorio et al. 2013). The texture and morphology of the organic matter on the nanometer- to micrometer-scale could be an additional key for understanding its formation and subsequent evolution, and should be investigated experimentally in more depth.

Because macromolecular organic matter formed in molecular clouds could be one of the main solid components of the outer part of the protoplanetary disks, its physical properties, such as viscoelasticity, could affect the aggregation efficiency of dust particles in protoplanetary disks (Bridges et al. 1996; Cuzzi and Weidenschilling 2006). Kudo et al. (2002) measured the temperature dependence of viscoelastic properties of molecular-cloud organic analogues prepared by mixing chemical reagents (as described in Kouchi et al. 2002), and concluded that the aggregation of dust coated with molecular-cloud organic matter would be promoted at 200-300 K. However, there have been no further studies of the viscoelastic properties of organic matter, especially organic matter synthesized under molecular cloud conditions.

In this study, we investigated the physical properties of photochemically synthesized MCORs and DCORs. These properties, such as texture, morphology, and viscoelasticity, have only been studied in a limited number of laboratory experiments (Jenniskens et al. 1993; Dworkin et al. 2001). The *Material and Method* section describes the experimental apparatus and protocol used to produce the MCORs and DCORs and details the analyses of organic residues. The *Results* section presents a morphological description of the MCORs and DCORs and reports their viscoelastic properties. Finally, we discuss the implications of the experimental results for the evolution of organic compounds through UV irradiation, the similarity between experimental residues and Solar System organic matter, and the role of organic compounds in grain aggregation in protoplanetary disks.

## 2. Materials and Methods

### 2.1. Experimental apparatus

We developed a new experimental apparatus for the photochemical synthesis of molecular cloud organic matter at Hokkaido University: PICACHU (Photochemistry in Interstellar Cloud for Astro-Chronicle in Hokkaido University). PICACHU is composed of a gas mixing line and a vacuum chamber ($\sim 10^{-7}$ Pa) connected to a closed-cycle helium refrigerator (Advanced Research System Inc. DE-204-AB with an 800-K heating interface) and deuterium UV lamps (Figure 1.a). The ultra-high vacuum is obtained in the main chamber by a turbomolecular pump with a capacity of 400 L s$^{-1}$ (for $N_2$) and a scroll pump.

Gas deposition and UV photon irradiation take place on three faces of an open rectangular cuboid (dimension of 26 × 28 × 40 in mm, Figure 1.a and b) cooled to ~10 K. Temperature is monitored by a silicon diode thermometer attached to the cold head of the closed-cycle helium refrigerator and a type E thermocouple attached to the 800-K heating interface. A mixture of gases, previously prepared in a gas mixing line, is introduced to the main chamber through three 2-cm-diameter capillary plates facing the three deposition surfaces of the cuboid. The use of capillary plates enables a homogeneous ice film to be deposited on the substrate. The gas deposition flux is adjusted by a variable leak valve. The UV photon sources are three water-cooled type 150-W deuterium lamps (Hamamatsu L1835) with $MgF_2$ windows pointed towards each deposition surface. The wavelength of the UV light from the lamps ranges from 110 to 170 nm, with peaks at 121.6, 125.4, and 160.8 nm. The photon flux of the lamps at the deposition surface is 0.5 to $1.5 \times 10^{14}$ photons $cm^{-2}$ $s^{-1}$. One deposition surface of the cuboid can be partially covered by a moving metallic plate attached to a linear motion feed-through in order to have different UV photon doses on a single surface.

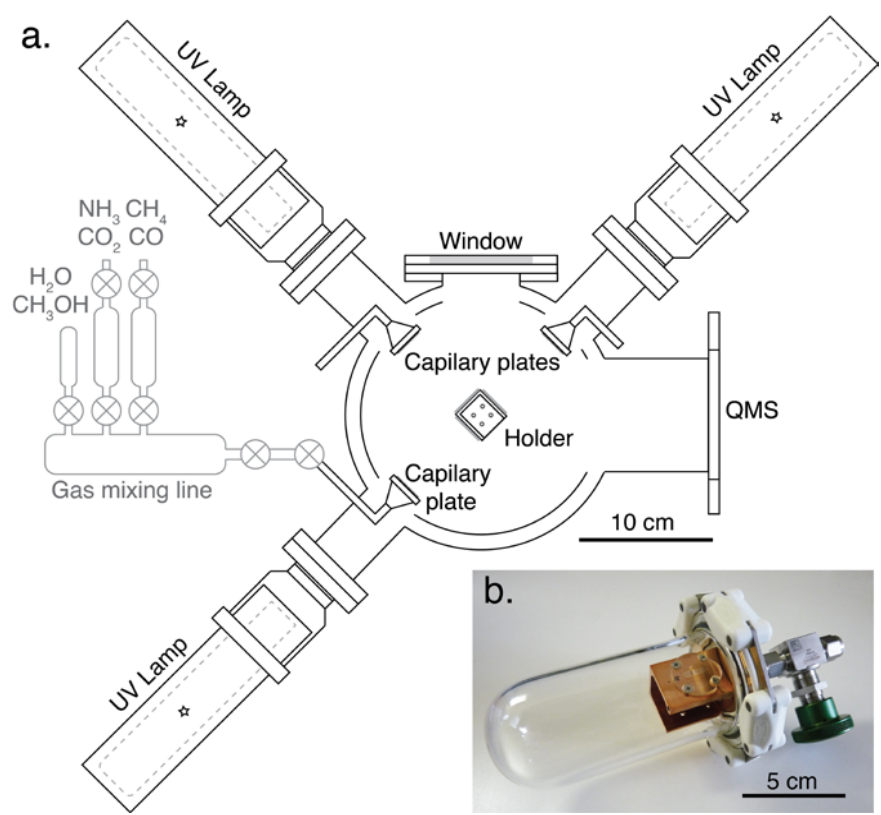

Figure 1. (a) Schematic view of the main chamber of PICACHU showing the positions of the capillary plates, the deuterium-lamps for UV irradiation, the sample holder, a quadrupole mass spectrometer (QMS) and an observation window. The metallic plate attached to a linear motion feed-through used to cover one of the deposition surfaces during UV irradiation is not represented. The gas mixing line is shown for indication, and is not represented to scale. (b) Picture of the hermetic container containing a cuboid holder with screwed corundum substrates fixed on the bottom of the container.

## 2.2. Experimental protocol

Experiments were performed following the protocol described below.

(1) The initial gas mixtures were prepared in the gas mixing line using $H_2O$, $CH_3OH$, and $NH_3$, three of the main components of interstellar gas and ice. Ultrapure water and liquid $CH_3OH$ from Wako Pure Chemical Industries, Ltd. with >99.8% purity (recommended for dioxin analysis) attached to the gas mixing line were repeatedly degassed by freezing and pumping to remove dissolved volatile species before use. The $H_2O$ and $CH_3OH$ vapors over the liquids were mixed with $NH_3$ from Sumitomo Seika Chemical Co. with >99.9995 % purity to prepare the desired mixtures. The gas mixture composition was monitored by a quadrupole mass spectrometer (Extrel Core Mass Spectrometer) connected to the main chamber. We did not observe any reaction products that may be formed by reactions between the mixed gases before their deposition on the cold surface. Water, $CH_3OH$, and $NH_3$ were mixed with a molar ratio of 2:1:1 for five experiments. Two experiments were conducted with ratios of 5:1:1 and 10:1:1 (Table 1).

(2) Simultaneous gas deposition and UV irradiation on the three deposition surfaces were carried out at ~10 K over 3 to 5 days (Table 1). The gas flux was set to 1.1-1.6 × $10^{14}$ molecules $cm^{-2} s^{-1}$ giving a gas molecule to photon ratio of 0.8 to 1.4. For comparison, we also performed blank experiments with UV irradiation but no gas or only $H_2O$, as well as blank experiments without UV irradiation and with $H_2O$ only or with $H_2O:CH_3OH:NH_3$ (2:1:1) (Table 1).

(3) After the desired duration, the gas deposition, UV irradiation, and cryostat were all turned off. The deposited ice warmed up to room temperature in about 7 hrs. The typical temperature increase rates were 4 K/min at 40 K, 2 K/min at 60 K, and 1 K/min at 130 K. At room temperature, whitish residues were found on the surface of the substrates in all experiments except the blank experiments. These residues were clearly derived from the ice formed at low temperature with UV irradiation, and are considered analogues of dense molecular cloud organics (MCORs).

(4) Further UV irradiation at room temperature was performed for 3 to 10 days on two of the three surfaces with organic residues. One of these irradiated surfaces was partially covered in order to obtain either duplicates or residues under different irradiation conditions (Table 1). Although, the temperature of grains in diffuse clouds is estimated to be about 20 K (Tielens, 2005), we choose to perform the irradiation of MCORs at room temperature in order to avoid the deposition of ice from the residual gas following previous studies (Jenniskens et al. 1993; Greenberg et al. 1995). Indeed, the deposition of a thick layer of ice, that is inevitable for long time experiments even with ultrahigh-vacuum conditions, would strongly reduced the efficiency of the UV-irradiation due to its optical thickness (Jenniskens et al. 1993). These re-irradiated samples are thus considered analogues of molecular cloud organics that have been reprocessed in the diffuse cloud (DCORs) or in circumstellar environments.

In our series of experiments (Table 1), we used different deposition substrates designed for different observational and analytical techniques (see section 2.3). Corundum disks or gold-coated cupper disks were used for most of the experiments. For the transmission electron microscope observations, MCORs and DCORs were produced directly on single crystalline Si grids, with 5- or 9-nm-thick amorphous Si films (SiMPore Inc. Pure Si TEM windows) in experiment P5 with $H_2O:CH_3OH:NH_3 = 2:1:1$.

Synthesized MCORs and DCORs were stored in a hermetic container with argon gas (Figure 1.b). No discernible change in the morphology of MCORs or DCORs was observed during storage (a few days to several weeks).

**Table 1**
Experimental Conditions

| Exp#[a] | Initial gases $H_2O:CH_3OH:NH_3$ | Low temp.[b] | Room temp.[c] | Deposition substrates |
|---|---|---|---|---|
| Blank 1 | 0:0:0 | 71 hrs | 0 hrs | cuboid holder |
| Blank 2 | 1:0:0 | 71 hrs | 0 hrs | cuboid holder |
| Blank 3 | 1:0:0 | 72 hrs / no UV | 0 hrs | cuboid holder |
| Blank 4 | 2:1:1 | 72 hrs / no UV | 0 hrs | cuboid holder |
| P1 | 2:1:1 | 72 hrs | 0 hrs | cuboid holder |
| P2 | 2:1:1 | 116 hrs | 91/65/39/0 hrs | 3 cor.[d] |
| P3 | 2:1:1 | 71 hrs | 65/39/0 hrs | 3 gold-coat.[e] |
| P4 | 2:1:1 | 71 hrs | 238/0 hrs | 3 cor.[d] |
| P5 | 2:1:1 | 71 hrs | 186/0 hrs | 2 TEM[f] + 1 cor.[d] |
| P6 | 5:1:1 | 96 hrs | 96/0 hrs | 2 cor.[d] + gold-coat.[e] |
| P7 | 10:1:1 | 96 hrs | 186/0 hrs | 2 cor.[d] + gold-coat.[e] |

**Notes.** [a]Run number, [b]duration of the low temperature deposition/irradiation phase, [c]duration of the room temperature UV irradiation, [d]corundum plate, [e]gold-coated cupper substrate, [f]transmission electron microscope grids fixed on a gold-coated cupper substrate.

## 2.3. Observation and analysis of organic residues

MCORs and DCORs were observed first by reflected light optical microscopy. In addition, field emission scanning electron microscopy (FE-SEM) was performed on the residues on gold-coated cupper substrates using the JEOL JSM 7000F at Hokkaido University, operated at 15kV (working distance of 5 mm).

Topographic images of MCORs and DCORs were obtained at Hokkaido University using a 3D laser-scanning microscope VK-X200 (KEYENCE Corp.) and an atomic force microscope (AFM) MFP-3D-BIO (Asylum research) operated in tapping mode with a silicon cantilever (Olympus AC240TS with a resonance frequency of 70 Hz and force constant of 2.7 N/m) or in contact mode with a silicon nitride cantilever (Olympus TR400PSA with a length of 200 μm and force constant of 0.02 N/m). AFM images were acquired at a resolution of 256 × 256 pixels (corresponding to 7.8 nm per pixel for a 2 μm × 2 μm image).

Transmission electron microscopy (TEM) was performed with a JEOL JEM-2100VL operated at 80 kV at Hokkaido University and with a JEOL JEM-2100F operated at 200 kV at Kyoto University. TEM observations of residues were made directly on TEM grids. Some residues on gold-coated copper disks were also observed after tearing them off from the disks using a Ge thin substrate.

The viscoelastic properties of MCORs and DCORs were measured at room temperature (23°C) by nano-indentation (Syed Asif et al. 1999) using a TriboIndenter (Hysitron, Inc.), with a 50 μm radius spherical indenter modulated at a frequency range of 1 to 250 Hz.

## 3. Experimental results

### 3.1. General aspect of the residues

With reflected light microscopy at low magnification, MCORs appear to form a network with voids (Figure 2.a and b). These voids are often round in shape and vary in size from a few tens to a few hundreds of micrometers (Figure 2.c and d). Experiment P7, in which the initial gas mixture was the richest in water ($H_2O:CH_3OH:NH_3=10:1:1$), appears to have produced the palest residues (Figure 2.d). The whitish MCORs (Figure 2.a) are soluble in water and methanol at room temperature, as was reported in previous studies (Greenberg 1983; Dworkin et al. 2004; Danger et al. 2013).

After room temperature UV irradiation, the color of the residues changed from whitish to yellow-brown (DCORs). The boundaries between MCORs and DCORs on samples partly covered during room temperature irradiation are clearly visible by naked eye or under the optical microscope (Figure 2.e and f). Roundish voids with no apparent organic deposit were also observed in DCORs that exhibited similar variations in size and distribution to MCORs.

Unlike MCORs, DCORs are not fully soluble in water and methanol, and even after a prolonged bath (several tens of minutes) in the solvents, part of the residue was still visible at the surface of the substrate. This is consistent with the recent report of de Marcellus et al. (2016) showing that molecular cloud organic analogues become partly insoluble in water after UV-irradiation.

It should be noted that the morphological and textural features of the MCORs and DCORs, described here and in the following section, are the same irrespective of the substrate used and experiment duration.

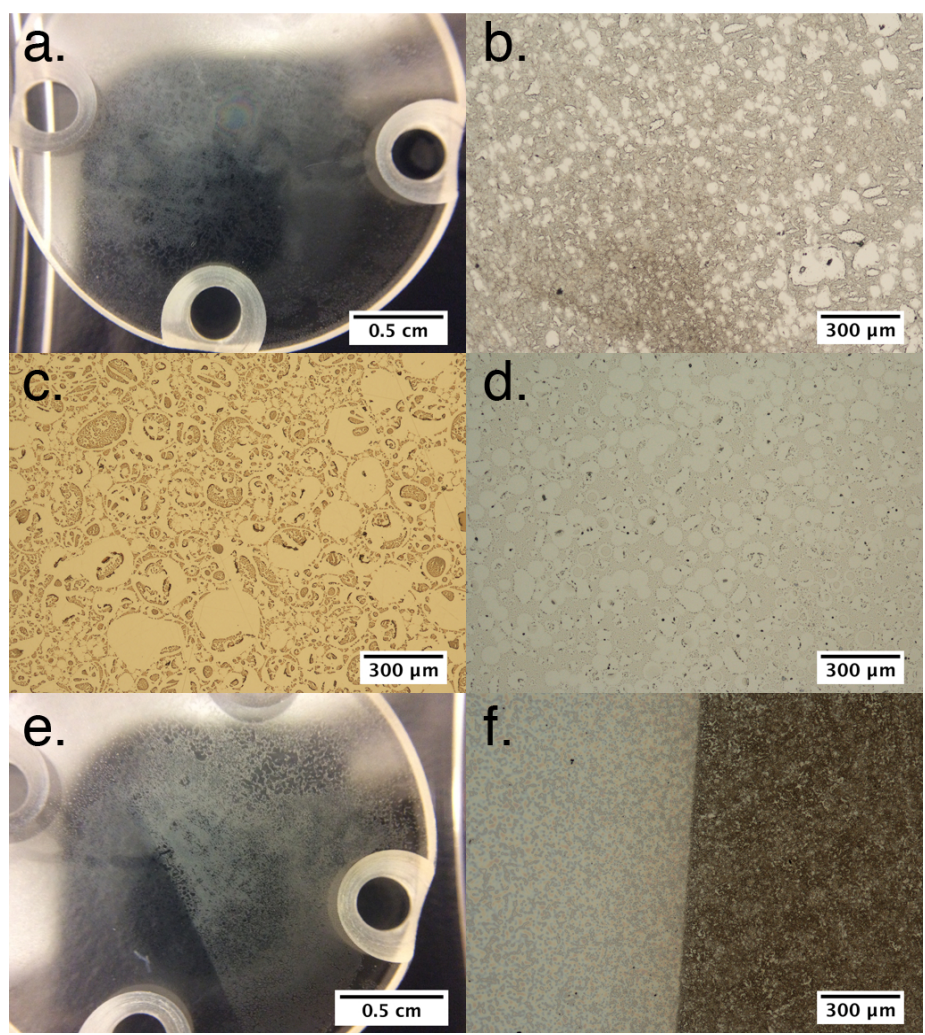

Figure 2. Images of MCORs and DCORs (a): large views of the MCOR on a corundum substrate (experiment P2) and (b, c and d): reflected-light microscope images of the MCOR from experiments P2, P6 and P7, respectively, (e and f): samples partly covered during room temperature UV irradiation showing MCORs and DCORs on the left and right parts of each image, respectively; large-scale view of the substrate in the experiment P2 (e) and reflected-light microscope image of the boundary between MCOR and DCOR in experiment P4 (f).

## 3.2. Morphologies and textural properties of MCORs and DCORs

The dark parts of the MCORs resemble an oily membrane stretched over the substrate, and show iridescent colors under reflected light at a high magnification (Figure 3.a, b). Elongated channels, deposited droplets and wrinkled-membrane-like structures are occasionally observed (Figure 3.b to f). The surface of the MCOR is smooth at the micrometer to sub-micrometer scale (Figure 3.c to h). The typical thickness of the deposit is about 100-300 nm (Figure 3.g and h). This is of the same order as the mean penetration depth of UV photons in pure $H_2O$, $CH_3OH$ or $NH_3$ ices (Cruz-Diaz et al. 2014) or organic compounds, in which the majority of energy deposited within the top ~0.1 μm of the surface (Greenberg et al. 1995; Bennett et al. 2013; Alata et al. 2014), allowing the majority of the residues to be processed with UV photons during irradiation at room temperature. The wrinkled-membrane-like structures can locally be as thick as 2-3 μm.

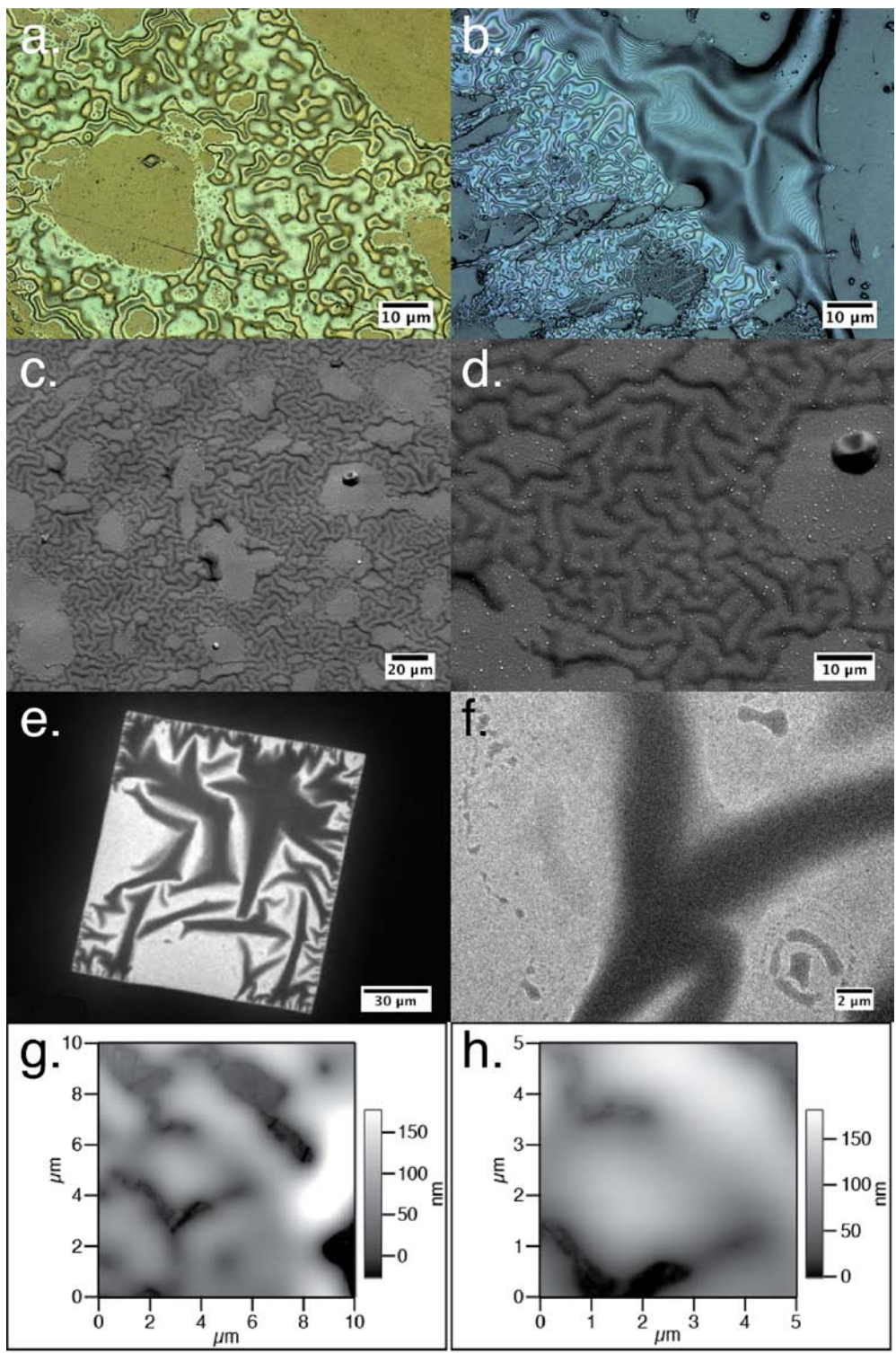

Figure 3. Representative images of MCORs taken by the laser-equipped microscope for experiment P3 (a) and P2 (b), by FE-SEM for experiment P3 (c and d), by TEM for experiment P5 (e and f), and by AFM for experiment P6 (g and h). MCOR appears as an oily membrane, partly wrinkled and very smooth at the micrometer to sub-micrometer scale.

In contrast, the DCORs show numerous rough patches irrespective of the starting gas composition or irradiation duration, which make these residues appear darker under the reflected-light optical microscope. Iridescent colors are more difficult to distinguish than they were in the MCOR (Figure 4.a and b). The thickness of the central part of the deposit network is a few hundred nanometers, while it reaches a few micrometers at the edges of the deposit (Figure 4.b, c, and d).

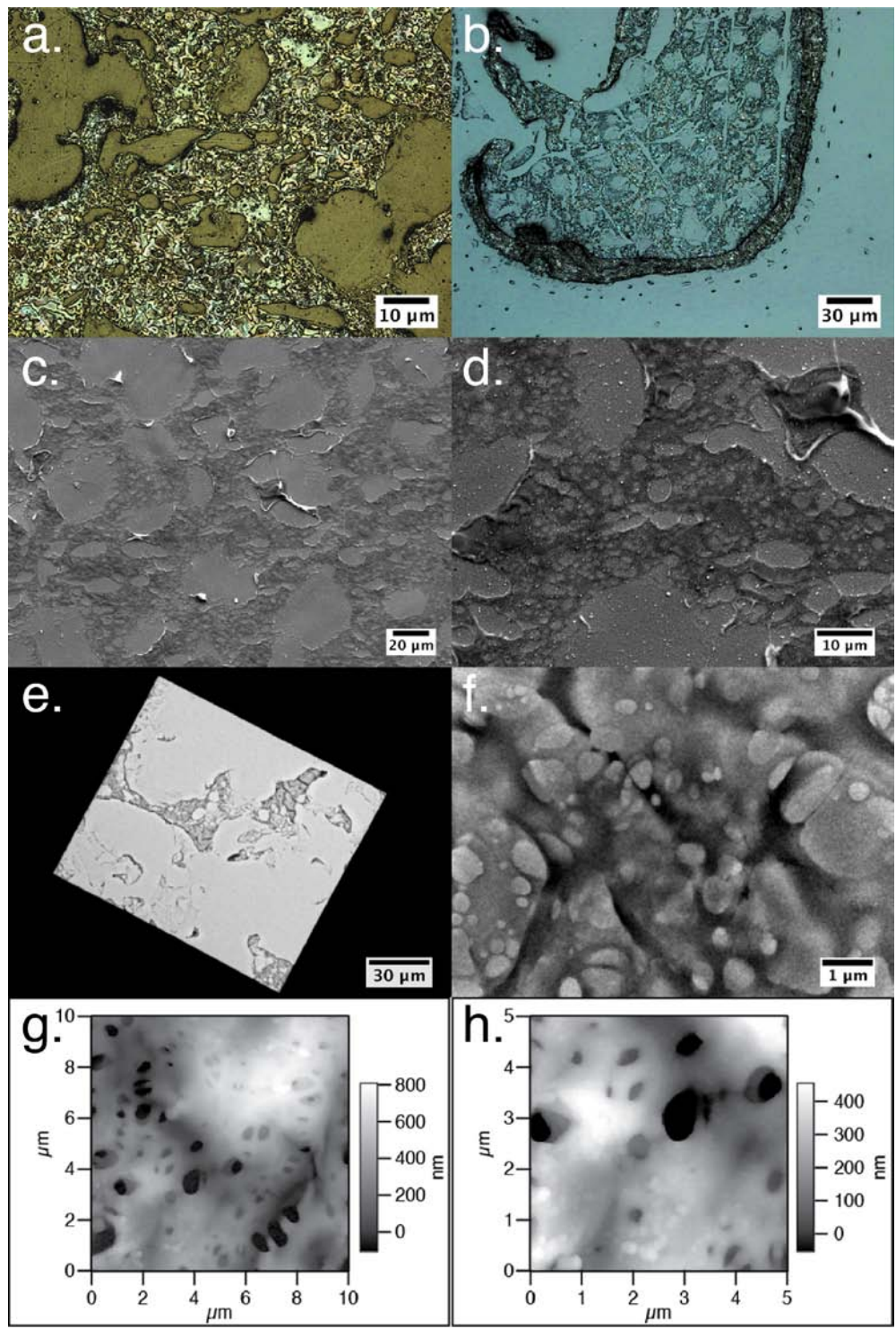

Figure 4. Representative images of DCORs taken by the laser-equipped microscope for experiments P3 (a) and P2 (b), by SEM for experiment P3 (c and d) by TEM for experiment P5 (e and f), and by AFM for experiment P2 (g and h). DCORs contain thickness variations, strong asperities, and porosity with micrometer, and sub-micrometer-sized pores.

Unlike the MCORs, the DCORs contain a large number of micrometer- and sub-micrometer-sized pores (Figure 4.c to h) and appear to have a rough texture at the sub-micrometer scale (Figure 4.h). In four 10×10 μm² AFM images, 298 pores were identified, and the pore width (the distance between the two walls of the pore) ranged from 60 nm to 2.7 μm. Even smaller pores, with widths as low as 10 nm, were observed by TEM (Figure 4.f). The apparent porosity at the surface of the DCORs in experiments P2 and P5, estimated from the ratio of the area of the pores over the total area covered by the residue in the four 10×10 μm² AFM images, is 18 ± 9 % (2σ, standard deviation). Pores are also visible throughout the deposits in the TEM images, which also exhibit a large number of superimposed pores (Figure 4.f). No clear difference in porosity or roughness was observed in DCORs that underwent different photon doses.

Another textural feature that distinguishes the DCORs from the MCORs is the presence of turned-up residue borders, likely formed by shrinkage of the residue due to UV irradiation (Figure 4). The DCORs may be less able to maintain contact with the substrate surface (i.e. have a lower wettability) than the MCORs. From topographic images of the residues obtained by AFM, we were able to estimate the contact angles between the organic residues and the corundum substrates. Assuming a spherical droplet shape, the contact angle $\theta$ can be approximated by the relation: $\theta = 2 \times \tan^{-1}(h/d)$, where $h$ and $2d$ are the height and width of the droplet, respectively. The $h$ and $d$ parameters were determined using height profiles at a right angle to the contact between the residues and the substrate (examples of profiles are shown in Figure 5). We obtained a contact angle of 12° ± 11° (2σ, standard deviation) from 20 profiles of the MCORs produced from gases with starting compositions of $H_2O:CH_3OH:NH_3$ = 2:1:1 and $H_2O:CH_3OH:NH_3$ = 5:1:1. The MCOR produced from a starting composition of $H_2O:CH_3OH:NH_3$ = 10:1:1 has higher contact angles (54° ± 11°, 2σ from 9 profiles). However, these high contact angles may not be representative of the whole sample because we were only able to obtain one AFM image from the sticky sample. The DCORs display higher contact angles, with an average of 77° ± 42° (2σ) from 11 profiles. Although the measurement of contact angles on the DCORs might not directly correspond to the wettability of the residue on corundum, the presence of the turned-up borders and the difference in contact angles imply a difference in the physical properties of MCORs and DCORs.

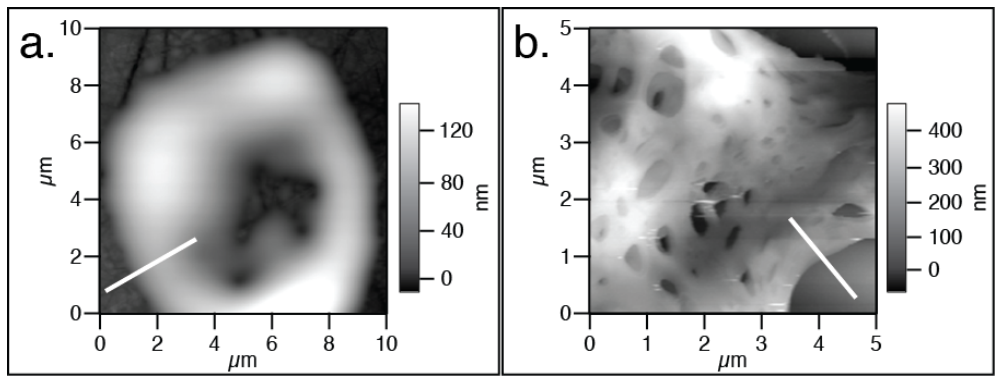

Figure 5. AFM images showing the contacts between (a) MCOR (experiment P6) and (b) DCOR (experiment P2) and the substrate (dark gray). The white lines show the locations of two height profiles (white lines) used for measuring contact angles.

At the micrometer to sub-micrometer scale, unique textures can also be observed in the DCORs. A thin part of the DCOR membrane contains numerous elongated or tubular features with diameters of a few tens to a few hundreds of nanometers, which constitute the borders of the pores formed during UV-irradiation, (Figure 6.a and b). Discrete particles, ≤10 nm in size (Figure 6.c), and spherical features with diameters of several tens of nanometers (Figure 6.d) can also be observed. Numerous discrete nanospheres, with diameters of 10-50 nm, form aggregates that seem to constitute the main part of the DCOR membrane surface, as revealed in the AFM (Figure 6.e and f) and FE-SEM images (Figures 6.g and h). These textures are not observed in the MCORs, indicating that their formation resulting from the UV irradiation.

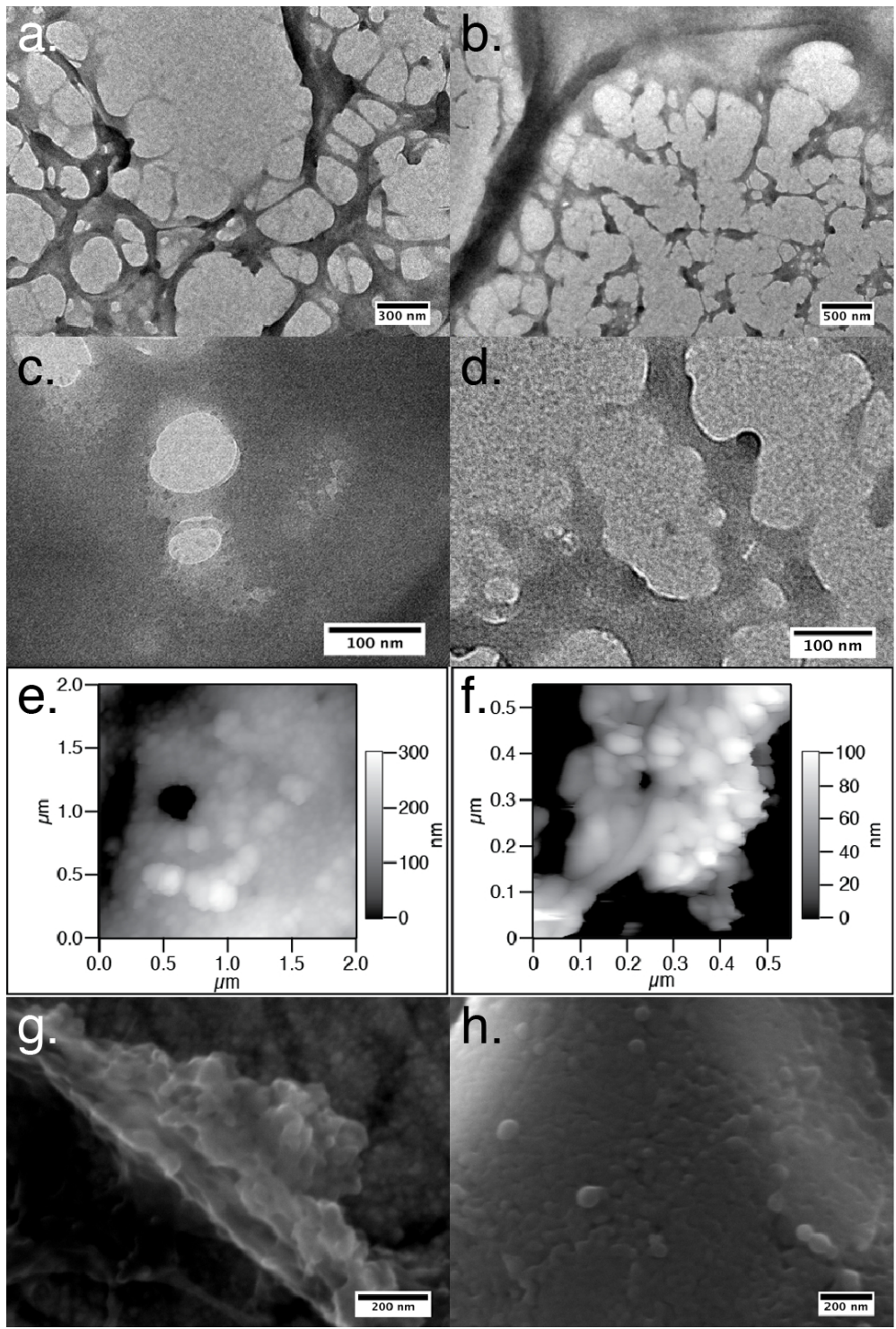

Figure 6. Nanostructure of the DCOR: TEM images of the DCOR in experiment P5, showing elongated features (a and b), discrete nanoparticles (c) and spherical tips (d); AFM images of experiments P2 (e) and P6 (f), and FE-SEM images of P6 (g and h), show that the residue is formed of aggregates of nanoparticles that are a few tens of nanometers in size.

## 3.2. Viscoelasticity of MCORs and DCORs

Oscillating stress $\sigma$ and strain $\varepsilon$ are obtained from the sinusoidal deformation of the organic residues by the spherical indenter in the nano-indentation technique. From these parameters, the storage modulus $E'$ and the loss modulus $E''$ can be derived:

$$\sigma/\varepsilon = E^* = E' + iE''$$

$E'$ and $E''$ can be written as

$$E' = \frac{k}{2}\sqrt{\frac{\pi}{A}}$$

$$E'' = \frac{\omega\eta}{2}\sqrt{\frac{\pi}{A}}$$

$$\frac{E''}{E'} = \frac{\omega\eta}{k} = \tan\delta,$$

where $k$ is the stiffness, $A$ is the cross-section of the indenter, $\omega$ is the angular frequency, $\eta$ is the viscosity, and $\delta$ is the phase shift between $\sigma$ and $\varepsilon$.

The storage modulus $E'$ and the loss modulus $E''$ are related to the elastic modulus and viscosity, respectively. Figure 7 shows the frequency dependence of $E'$, $E''$ and the dissipation factor (*tan δ*) of the MCOR and DCOR of experiment P4 (Table 1) measured at 23°C. At low frequency (from 1 to 10 Hz) for the MCOR, $E''$ is almost constant (~200 MPa) and larger than $E'$, indicating the viscous character of the residue. This viscous nature is consistent with the TEM observations of bridge-like structures between MCOR films that were torn apart in the TEM grid (Figure 8). The loss modulus $E''$ decreases to 30-70 MPa at 10-70 Hz and increases to 200-300 MPa from 100 Hz, while the storage modulus $E'$ is larger than $E''$ at the frequency of >10 Hz and increases sharply with frequency. The MCOR thus changes from a viscous to an elastic material when the frequency of the deformation is higher than 10 Hz.

For the DCOR, $E'$ and $E''$ are constant over the entire frequency range, at 100 MPa and 20 MPa, respectively. The small but constant values of $E'$ clearly show that the DCOR is more fragile than the MCOR, which is consistent with the cracked and porous nature of the residue as observed with Laser microscope, TEM, and AFM (Figure 4).

No discernible change was observed in the morphology of MCORs and DCORs after removal from the vacuum chamber. This is consistent with the non-significant effect of air exposure on the infrared spectra of organic residues reported by Muñoz Caro et al. (2004). Nonetheless, water vapor absorption onto the samples cannot be completely ruled out during storage and measurement, especially for the MCORs. Water adsorption may result in deliquescence of the samples (Mikhailov et al. 2009), and the viscosity parameter (*E″*) measured in the present study might therefore be a lower limit for the MCOR.

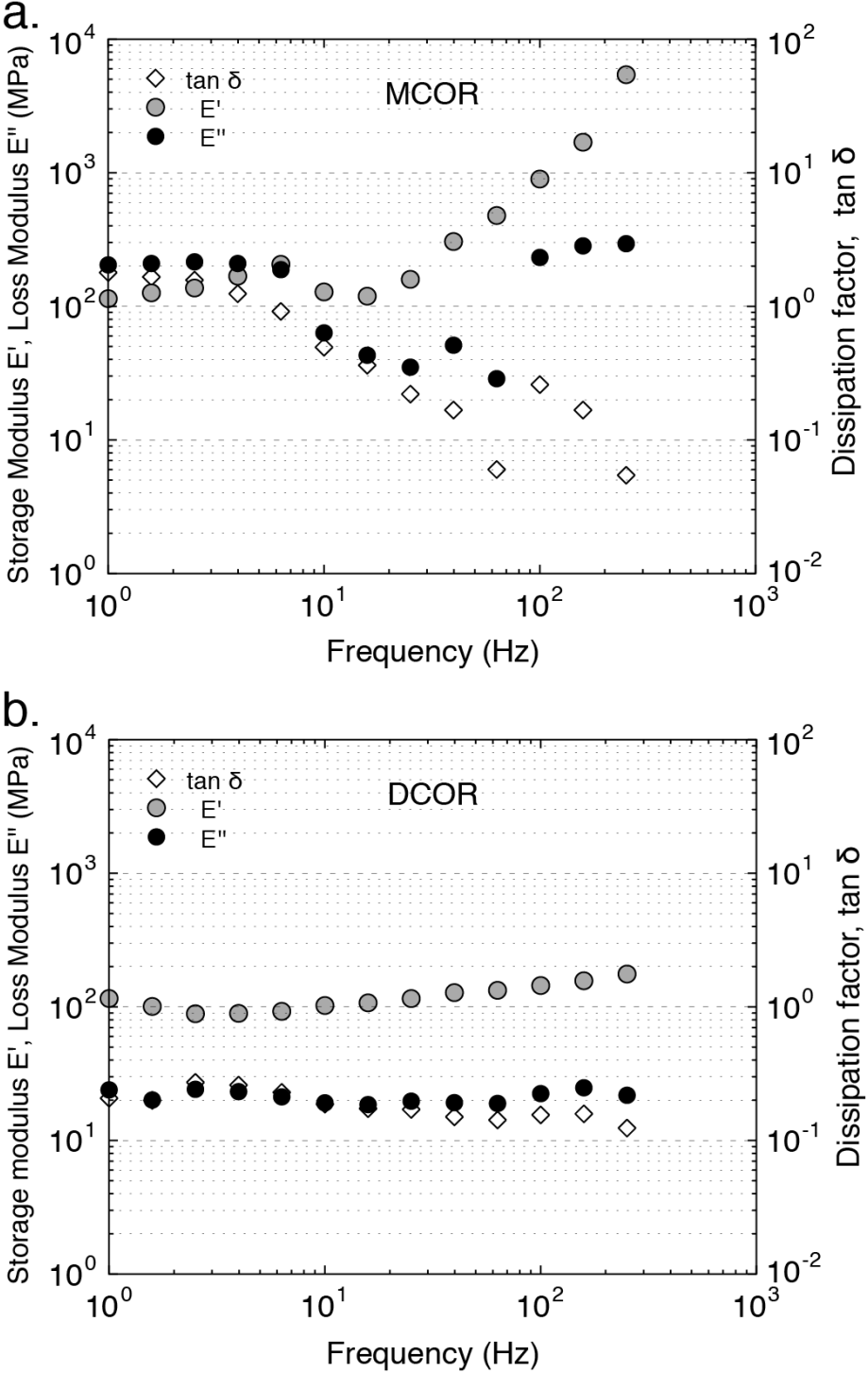

Figure 7. Frequency dependence of the storage modulus *E'*, the loss modulus *E″* and the dissipation factor *tan δ* of the (a) MCOR and the (b) DCOR in experiment P4 measured at 23°C.

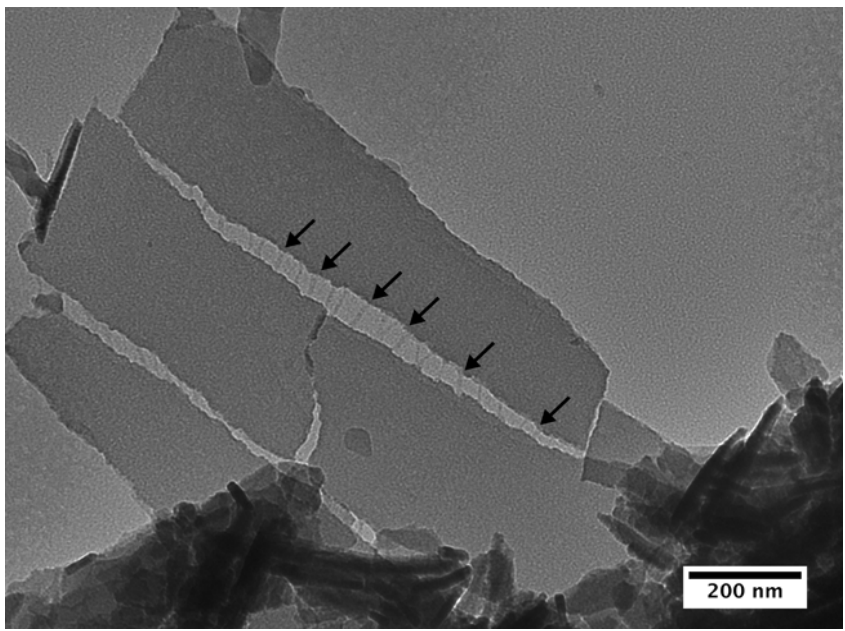

Figure 8. TEM observation of bridge-like structures between two MCOR films (shown with arrows).

## 4. Discussion and astronomical implications

### 4.1. Rapid evolution of the dense molecular cloud organics under UV irradiation

Previous experiments have reported the darkening and warping of residues after UV irradiation, associated with the loss of H, N, O hetero-elements (carbonization) and functional groups, and a decrease in density (Jenniskens et al. 1993; Greenberg et al. 1995; de Marcellus et al. 2016). In our experiments, the molecular cloud analogues irradiated by UV photons at room temperature (DCORs) show systematic and clear differences in color, morphology, porosity, and viscoelastic properties compared to the molecular cloud analogues (MCORs) (Figures 3-8). The strong increases in roughness, porosity, and shrinkage of the residues after UV irradiation at room temperature (Figures 3 and 4) argues for their evaporation during UV irradiation, and the porosity increase might also explain the lower bulk density of the irradiated residues compared to the non-irradiated residues reported by Jenniskens et al. (1993). The contact angle on the corundum substrate and the viscoelastic properties of the residues are also clearly modified by the photo-process (Figures 5 and 8).

In the diffuse cloud, the photon flux in the range of 210 nm > λ > 90 nm is ~$10^8$ photon $cm^{-2}$ $s^{-1}$ (Moore et al. 2001). Hence, the photon doses for the DCORs in our experiments are equivalent to a residence duration of $10^3$-$10^4$ yrs in the diffuse clouds. This duration is much shorter than the lifetime of the molecular cloud ($10^6$-$10^7$ yrs, Moore et al. 2001) and molecular cloud organic-coated grains would therefore easily have been subjected to a photon dose equivalent to the experimental ones (Greenberg et al. 2002). The present results suggest that the alteration of organic matter by UV photons in the diffuse cloud occurs easily and causes drastic changes to its morphological and physical properties.

We also note that our experimental irradiation conditions might also be consistent with irradiation in the outermost part of the protoplanetary disks (Carr and Najita 2008, 2014). Taking into account the lifetime of the disk ($10^6$-$10^7$ yrs) and the particle movement due to gas drag, gravitational settling, and turbulent diffusion, a µm-sized particle would receive around $5 \times 10^{12}$ photons (or a photon dose of ~$1.6 \times 10^{20}$ photon cm$^{-2}$ for a spherical particle)(Ciesla and Sandford 2012). The photon doses in our experiments are 2–16 times smaller than this dose, indicating that the morphology and physico-chemical properties of the organic mantle of dust particles could also have been modified in the outer protoplanetary disks.

**4.2. Nano-structure of DCORs and their relevance to organic compounds in chondrites and IDPs**

One of the most striking observations of the DCORs is the presence of aggregates of nanoparticles (Figure 6) similar in size (10-50 nm) to the smallest nanoglobules observed in chondrites and IDPs. In the organic matter of carbonaceous chondrites (CI, CO, CM, CR and ungrouped Tagish Lake), spherical particles with diameters of <20 nm to a few micrometers have been found forming clusters and intergrowths (Garvie et al. 2008; De Gregorio et al. 2013). The smallest nanoglobules are observed in IOM from the least altered, and thus better preserved, chondrites (De Gregorio et al. 2013). Small amorphous nanoglobules are also found in the chondritic porous IDPs (Matrajt et al. 2012), which show no evidence for thermal or aqueous processes (Ishii et al. 2008), and in the Wild 2 cometary particles (De Gregorio et al. 2010).

Although different processes, including parent body aqueous alteration, have been proposed for the formation of spherical nanoparticles and nanoglobules, we propose that UV irradiation of molecular cloud organic compounds in the diffuse cloud or in the outer protoplanetary disk could also be a possible mechanism for globule formation, especially for small nanoglobules in samples that were altered the least on their parent bodies. This is in line with the work of Nakamura-Messenger et al. (2006), suggesting that hollow nanoglobules might be remnants of UV-irradiated organic mantles around less complex organic compounds or dust grains protected from UV that have been lost at a later stage forming hollow globules.

Deuterium-enriched nanospherules in CR chondrites and in chondritic porous IDPs (Matrajt et al. 2012; Busemann et al. 2009; Hashiguchi et al. 2013), both of which experienced the least degree of parent body alteration, might be good candidates for having been formed under UV-irradiation of molecular cloud organics. Extreme D-enrichments in methanol and ammonia have been observed towards molecular clouds ($CH_2DOH/CH_3OH$ of up to 0.9 (Parise et al. 2002, 2004) and $NH_2D/NH_3$ of up to 0.3 (Ceccarelli 2002; Hatchell 2003)). Because DCOR is likely formed by UV irradiation of ices, including $CH_2DOH$ and $NH_2D$, in molecular clouds and by further UV irradiation of MCOR in diffuse cloud or outer protoplanetary disks, it is also possible that D-enriched DCORs containing nanospherules could form. The fact that D-enriched amino acids are formed by UV irradiation of $H_2O$-CO-$CH_2DOH$-$NH_3$ ice (Oba et al. 2016) also shows that D-rich complex organic compounds can be produced via this pathway.

**4.3. Carbon-coated grain aggregation in the proto-planetary disk**

In protoplanetary disks, dust and ice grains experience collisions with a typical maximum impact velocity of ~50 m s$^{-1}$ (e.g. Johansen et al. 2014). Depending on the physical properties of the grains and their collision velocity, these impacts can result either in merging or disruption of the grains. In particular, the maximum impact velocity for the sticking of grain aggregates was shown to be inversely proportional to the cube root of the storage modulus $E'$ of the grains ($\propto 1/\sqrt[3]{E'}$, Wada et al. 2013). For aggregates composed of 0.1 μm-sized particles of ice with a storage modulus of 7 GPa, Wada et al. (2013) estimated a maximal impact velocity of 80 m s$^{-1}$ for sticking. This value is comparable to the typical impact velocity of grains in the disk, implying that a fraction of the grains could merge and grow into aggregates.

In the warm region of protoplanetary disks, icy components evaporate and dust aggregates are mainly composed of silicate particles. The higher storage modulus of silicates compared to ice lowers the maximal impact velocity for sticking (8 m s$^{-1}$ for 0.1 μm-size silicate particles, Wada et al. 2013) and reduces the efficiency of aggregate growth, which would stall at the cm-size (Johansen et al. 2014). However, if the grains were coated with DCOR-like organic compounds, the storage modulus of the particles could decrease to down to ~ 100 MPa (Figure 7). This would increase the maximum impact velocity for sticking to ~300 m s$^{-1}$ according to the empirical formula of Wada et al. (2013), and grain aggregates in the warmer region of the disk could then exceed the cm-size and grow to form planetesimals.

While organic coating on dust grains favors the aggregation of grains during impact, the viscoelastic properties of the organic components would affect the internal structure and brittleness of the dust aggregates and could reduce the efficiency of the dust aggregation. The viscoelastic behavior of the organic-coating of particles in aggregates would allow for deformation that would gradually enlarge the contact area between the particles, strengthening the particle connections in the same way as sintering. Sintering has been known to affect the dynamic properties of dust aggregates by reducing their ability to accommodate shock deformations and dissipate energy (Sirono 1999). Similarly, the organic-coating on dust grains could increase the brittleness of the grain aggregates and favor the disintegration of the aggregate by repetitive collisions.

Thus it is clear that the viscoelastic properties of the organic-coating on dust grains strongly affect the growth (or fragmentation) of dust aggregates. Further investigations, including the testing of the temperature dependence of the viscoelastic properties of the experimental analogues, would help us to better constrain the efficiency of collisional growth in the protoplanetary disk.

## 5. Conclusions

In order to investigate their physical properties, organic residues were synthesized under dense molecular cloud-like and diffuse cloud-like conditions using the new PICACHU experimental apparatus. We found that the morphology and physical properties of the organic residues evolve rapidly with UV irradiation at room temperature. Over a timescale equivalent to short residence durations in diffuse clouds ($10^3$-$10^4$ years) or protoplanetary disks ($< 10^6$ years), irradiated organic residues become highly porous and more rigid, and form amorphous nanoparticles that resemble organic compounds found in least-altered chondrites, chondritic porous IDPs and cometary samples. Interestingly, UV-irradiated organic residues could also have chemical similarities with the insoluble organic matter in chondrites as suggested by a recent report of their infrared spectra (de Marcellus et al. 2016). The viscoelastic properties of the organic residues show that the presence of organic coating on grains in protoplanetary disks could not only enhance the efficiency of dust aggregation due to its small storage modulus, but it might also reduce the growth efficiency as the viscoelastic nature of organic coating could strengthen the grain-grain connections and thus increase the brittleness of the dust aggregates. UV-irradiation in the diffuse clouds or in the outermost part of the protoplanetary disks therefore appears to have a strong influence on the evolution of the precursor organic compounds of protoplanetary disks.

**Acknowledgments**

Laser and atomic force microscope observations were carried out at the OPEN FACILITY, Hokkaido University Sousei Hall. Naoya Sakamoto and Tomokazu Yoshizawa are thanked for their technical assistance and helpful suggestions. An anonymous reviewer is thanked for constructive comments that helped to improve the paper. We are also grateful to Associate Editor Steve Federman for careful editing. This work was supported by the grant-in-aid for Scientific Research on Innovative Areas "Evolution of molecules in space from interstellar clouds to proto-planetary nebulae" supported by the Ministry of Education, Culture, Sports, Science & Technology, Japan (Grant numbers 25108002, 25108003, 25108006).